\documentclass[aps,pre,reprint,showpacs]{revtex4-1}
\usepackage{graphicx}
\usepackage{dcolumn}
\usepackage{amsmath}
\usepackage{amssymb}
\usepackage{color}
\usepackage{bm}
\newcommand{\ave}[1]{\ensuremath{\left\langle#1\right\rangle} }

\begin{document}
\preprint{HEP/123-qed}
\title{Macroscopically measurable force induced by \\ temperature discontinuities at solid-gas interfaces}
\author{Masato Itami}
\affiliation{Department of Physics, Kyoto University,
Kyoto 606-8502, Japan}
\author{Shin-ichi Sasa }
\affiliation{Department of Physics, Kyoto University,
Kyoto 606-8502, Japan}
\date{\today}

\begin{abstract}
We consider a freely movable solid that separates a long tube into two regions, each of which is filled with a dilute gas.
The gases in each region are initially prepared at the same pressure but different temperatures.
Under the assumption that the pressure and temperatures of gas particles before colliding with the solid are kept constant over time, we show that temperature gaps appearing on the solid surface generate a force.
We provide a quantitative estimation of the force, which turns out to be large enough to be observed by a macroscopic measurement.
\end{abstract}

% 05.70.Ln, Nonequilibrium and irreversible thermodynamics
% 05.40.-a, Fluctuation phenomena, random processes, noise, and Brownian motion
% 05.60.-k, Transport processes

\maketitle

\section{Introduction}
% macroscopic non-equilibrium dynamics and beyond
Macroscopic nonequilibrium phenomena are described by evolution equations for slow modes associated with conservation laws and symmetry breaking \cite{Chaikin}.
The equations for standard liquids and gases are well established as the hydrodynamic equations of mass, momentum, and energy density fields, and heat conduction and sound propagation in solids are also well established \cite{Landau}. 
The validity of the description was carefully investigated in small-scale experiments \cite{Granick}, which suggested that the behavior near a solid wall shows deviations from calculation results based on the standard hydrodynamic equations.
Furthermore, a stimulating prediction that a liquid droplet is nucleated in a sheared solid may be another example that is not described by the established continuum equations \cite{Biroli}.

% interfaces and motivation 

In these examples, nontrivial phenomena occur at the interface between a solid and a fluid.
Indeed, the description of behavior near the interface has not been established, because its characteristic length scale is too small for macroscopic phenomenological descriptions.
Although imposing appropriate boundary conditions at the interface for macroscopic equations often gives a good description, there are cases where the assumptions of the boundary conditions should be seriously considered.
In this paper, we study a phenomenon associated with temperature gaps at the interfaces between a solid and gases. 

%% idealized situation and MDD

\begin{figure}
\centering
\includegraphics[width=0.90\linewidth]{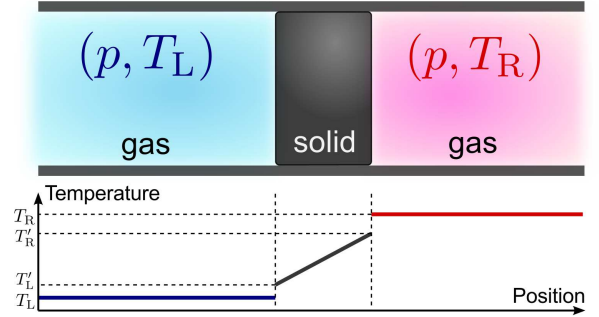}
\caption{(Color online) Schematic illustration of the experimental systems. The solid separates two gases at the same pressure but different temperatures. The pressures and temperatures are kept constant over time. Temperature gaps appear on the solid surface. }
\label{fig1}
\end{figure}

In order to demonstrate our findings clearly, we employ the special-purpose systems shown in Fig.~\ref{fig1}.
A solid (say, silicon), which consists of many atoms, is placed in a long tube of cross-sectional area $S$.
Dilute gases (say, helium) at the same pressure $p$ but different temperatures $T_{\rm L}$ and $T_{\rm R}$ are contained in the left and right regions, respectively, at an initial time.
The gases are well approximated by ideal gases and cannot mix with each other because the solid acts as a separating wall.
It is assumed that the pressure and temperatures of gas particles before colliding with the solid are kept constant over time.
In this setup, despite the equal pressure, momentum flows from one gas to the other owing to the energy transfer from the hot side to the cold side.
Recently, a phenomenological mechanism for the emergence of a force from such cross-coupling has been proposed in Refs.~\cite{Fruleux, Kawai}.
In this paper, we provide a quantitative estimation of the force acting on the solid on the basis of a microscopic description of the system under some assumptions.
An important finding is that the force is determined by the temperature gaps at the interface of the solid and gases.

%% our result 
More precisely, we denote by $T_{\rm L}'$ and $T_{\rm R}'$ the kinetic temperatures of the solid particles at the left and right ends, respectively, which are different from $T_{\rm L}$ and $T_{\rm R}$ in general.
Such a temperature discontinuity at an interface has been measured in experiments \cite{Swartz,Ward,Cahill}.
For the mass of gas particles $m_{\rm G}$ and the mass of solid particles $m$, we define $\epsilon\equiv \sqrt{m_{\rm G}/m}$, which is assumed to be
small.
We then show that the temperature gaps $T_{\rm L}' - T_{\rm L}$ and $T_{\rm R}'-T_{\rm R}$ generate the force $F_{\rm gap}$ given by
\begin{equation}
F_{\rm gap} = \epsilon^2 pS \left(
\frac{T_{\rm L}' - T_{\rm L}}{T_{\rm L}} +
\frac{T_{\rm R}-T_{\rm R}'}{T_{\rm R}}\right).
\label{eq:F_gap}
\end{equation}
By assuming Fourier's law in the solid, we can estimate the temperature gaps in terms of the thermal conductivity of the solid.
Surprisingly, the result shows that $F_{\rm gap}$ takes a macroscopically measurable value. 
It should be noted that steady-state motion of the solid is observed because the force $F_{\rm gap}$ may be balanced with a friction force induced by collision with gas particles.

%% Road map 

In the argument below, we describe the microscopic model that we employ.
We then derive the aforementioned result. 
Finally, we discuss the possibility of experimental realization of the phenomenon in laboratories.
Throughout the paper, $\beta$ represents the inverse temperature and $k_{\rm B}$ the Boltzmann constant.
The subscripts or superscripts ${\rm L}$ and ${\rm R}$ represent quantities on the left and right sides, respectively.

\section{Model}
\begin{figure}
 \centering
 \includegraphics[width=0.90\linewidth]{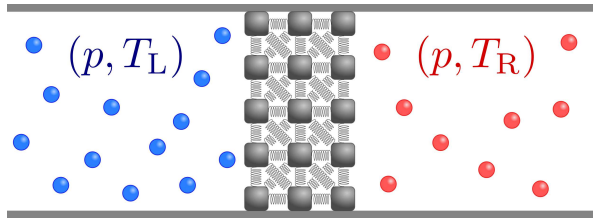}
 \caption{(Color online) Schematic illustration of a cross section of our model in three dimensions. The solid particles are connected by a spring. Gas particles collide with the solid particles according to a Poisson process.}
\label{fig2}
\end{figure}

%% micro-state
We provide a three-dimensional mechanical description of the solid in Fig.~\ref{fig2}.
We take the $x$ axis along the axial direction of the tube.
We assume that the solid consists of $N\times M$ particles of mass $m$, where $N$ and $M$ are the number of particles along the $x$ direction and in a plane perpendicular to the $x$ axis, respectively.
A collection of the positions and momenta of $N\times M$ solid particles, which we distinguish by subscripts $i$ and $j$ ($ 1 \le i \le N$, $1 \le j \le M)$, are denoted by $\Gamma = (\bm{r}_{1,1},\dots ,\bm{r}_{N,M};\bm{p}_{1,1},\dots ,\bm{p}_{N,M})$, which gives the microscopic state of the solid.
The $x$ components of $\bm{r}_{i,j}$ and $\bm{p}_{i,j}$ are denoted by $x_{i,j}$ and $p_{i,j}$, respectively, and the corresponding velocity is given by $v_{i,j} \equiv p_{i,j} / m$. 
The position and velocity of the center of mass of the solid in the $x$ direction are denoted by $X$ and $V$, respectively.

%% Hamiltonian

The Hamiltonian of the solid, $H(\Gamma)$, is given by 
\begin{equation}
H (\Gamma ) =\sum_{i,j}
\left[ \frac{\left\vert \bm{p}_{i,j} \right\vert^2}{2m}
+ U_{\rm w}(\bm{r}_{i,j}) \right]
+ \hspace{-3mm} \sum_{\ave{i,j;i',j'}} \hspace{-2mm}
U_{\rm int}(\bm{r}_{i,j},\bm{r}_{i',j'}),
\label{eq:Hamiltonian}
\end{equation}
where $\ave{i,j;i',j'}$ represents the nearest- and second-nearest-neighbor pair of solid particles.
$U_{\rm int}$ is the interaction potential between two solid particles, and $U_{\rm w}$ is the potential between a solid particle and the tube wall.
The tube wall is assumed to be frictionless, and $U_{\rm w}(\bm{r}_{i,j})$ does not depend on $x_{i,j}$.
The motion of solid particles except for left and right ends is described by the Hamiltonian equations.

%% gases model (assumption)

Next, we provide an effective description of the gases. 
We focus on the gas on the left side; the gas on the right side can be described similarly.
Employing a dilute gas that consists of particles of mass $m_{\rm G}$, we may assume that the characteristic time of the dissipation process inside each gas is much longer than the time during which we observe the steady-state motion of the solid.
Therefore, gas particles that have yet to collide with the solid are in equilibrium at the temperature $T_{\rm L}$, the pressure $p$, and the number density $n_{\rm L}=p\beta_{\rm L}$.
We also assume that gas particles elastically and instantaneously collide with the solid only once.
More precisely, the instantaneous collision means that the characteristic time of the solid-gas interaction is much shorter than the relaxation time of all solid particles.
It should be noted that this situation is completely different from a case where the solid is effectively described as a wall with one degree of freedom.
For convenience, we make a list of characteristic time scales in Table~\ref{tab:time_scales}.
Our assumption means that $\tau_{1} \gg \tau_{2} > \tau_{3} \gg \tau_{4}$.
Furthermore, for simplicity, we assume that the tangent plane at the collision point is perpendicular to the $x$ axis, so that the $x$ component of the velocity of each solid particle at both ends is independent of the other components at the collisions.

\begin{table}
\caption{\label{tab:time_scales} Characteristic time scales.}
\begin{ruledtabular}
\begin{tabular}{cl}
symbol & definition\\
\hline
$\tau_{1}$ & relaxation time of the dissipation process inside gas\\
$\tau_{2}$ & relaxation time of $V$\\
$\tau_{3}$ & relaxation time of all solid particles\\
$\tau_{4}$ & solid-gas interaction time\\
\end{tabular}
\end{ruledtabular}
\end{table}

%% collision rate 

For this setup, the interaction between the solid and the gas on the left side can be described by random collisions with the collision rate $\lambda_{\rm L}(v_{\rm G},v_{1,j})$ per unit area for the gas particle velocity $v_{\rm G}$ and the solid particle velocity $v_{1,j}$.
The collision rate is explicitly written as
\begin{equation}
\lambda_{\rm L}(v_{\rm G},v_{1,j}) = n_{\rm L} (v_{\rm G}-v_{1,j})
\theta (v_{\rm G}-v_{1,j})f_{\rm eq}^{\rm L}(v_{\rm G}),
\label{eq:lambda_L}
\end{equation}
where $\theta$ represents the Heaviside step function and $f_{\rm eq}^{\rm L}(v_{\rm G}) =\sqrt{\beta_{\rm L} m_{\rm G}/2\pi} \exp \left( -\beta_{\rm L} m_{\rm G}v_{\rm G}^2 /2 \right)$ is the Maxwell--Boltzmann distribution.

%% evolution equation

The $l$-th collision time of a gas particle and the $j$-th solid particle at the left end is determined according to the Poisson process.
Suppose that a gas particle with a velocity in the $x$ direction $v_{{\rm G},j}^{{\rm L},l}$ collides with the $j$-th solid particle at $t=t_{j}^{{\rm L},l}$. 
The equation of motion for the $j$-th solid particle in the $x$ direction is written as
\begin{equation}
\frac{dp_{1,j}}{dt} = -\frac{\partial H(\Gamma)}{\partial x_{1,j}} + F_{{\rm L},j},
\label{eq:EOM_L}
\end{equation}
with
\begin{equation}
F_{{\rm L},j} = \sum_l
I\left( v_{{\rm G},j}^{{\rm L},l}, \tilde{v}_{1,j}\right)
\delta\left( t-t_{j}^{{\rm L},l}\right) ,
\label{eq:force_col}
\end{equation}
where $F_{{\rm L},j}$ is the force exerted by the elastic collisions of the gas particles, $I(v_{\rm G},v)=2m_{\rm G} m(v_{\rm G}-v)/(m_{\rm G}+m)$ the impulse of the collision, and $\tilde{v}_{1,j}(t)\equiv \lim_{t' \nearrow t} v_{1,j}(t')$ the velocity just before the collision when $t=t_{j}^{{\rm L},l}$.
Similarly, the collision rate of the gas on the right side is given by $\lambda_{\rm R}(v_{\rm G},v_{N,j}) = n_{\rm R} (v_{N,j}-v_{\rm G}) \theta (v_{N,j}-v_{\rm G})f_{\rm eq}^{\rm R}(v_{\rm G})$, and the equation of motion for the solid particles at the right end is determined as well.

%% NESS and notations

The Hamiltonian equations in combination with the Poisson processes yield a unique steady state.
The expectation value in the steady state is denoted by $\ave{\, \cdot\, }$.
We then have  $\ave{v_{i,j}}=\ave{V}$ for any $i$ and $j$, and we assume that the statistical properties in the steady state are homogeneous in the vertical direction.

\section{Analysis}
%% Equation of motion of the center of mass
We consider the equation of motion for the center of mass.
From the law of action and reaction, the equation in the $x$ direction is written as $mNMdV/dt = \sum_{j=1}^{M} \left[ F_{{\rm L},j} + F_{{\rm R},j} \right]$.
Thus, the total force acting on the solid in the $x$ direction is generated by the elastic collisions of the gas particles, where the collision rate depends on the velocity of the solid particles.
Because $d\ave{V}/dt =0$, we have the force balance equation in the steady state as $\sum_{j=1}^{M} \left[ \ave{F_{{\rm L},j}} + \ave{F_{{\rm R},j}} \right] =0$.
Here, we note that 
\begin{equation}
\ave{F_{{\rm L},j}} = \ave{\int dv_{\rm G}\lambda_{\rm L}(v_{\rm G},v_{1,j})I(v_{\rm G},v_{1,j})}\frac{S}{M}.
\end{equation}
We then expand the total force $\sum_{j=1}^{M} \left[ \ave{F_{{\rm L},j}} + \ave{F_{{\rm R},j}} \right]$ in $\epsilon\equiv\sqrt{m_{\rm G}/m}$. 
Defining $\gamma_{\rm L}\equiv \epsilon n_{\rm L}\sqrt{8mk_{\rm B}T_{\rm L}/\pi}$ and $\gamma_{\rm R}\equiv \epsilon n_{\rm R}\sqrt{8mk_{\rm B}T_{\rm L}/\pi}$, we obtain
\begin{align}
\sum_{j=1}^{M}\big[ &-\gamma_{\rm L}\ave{v_{1,j}}-\gamma_{\rm R}\ave{v_{N,j}}
+\epsilon^2 p \beta_{\rm L}m \ave{v_{1,j}^2}\notag \\ 
&-\epsilon^2 p\beta_{\rm R}m \ave{v_{N,j}^2} \big] S/M +O(\epsilon^3)=0.
\label{eq:balance}
\end{align}
See Appendix \ref{sec:appendix} for the derivation.
The first and second terms in (\ref{eq:balance}) are interpreted as the friction force that originates from the change in the collision rate due to the motion of the solid particles.
The terms proportional to $\epsilon^2$ in (\ref{eq:balance}) are expressed in the form (\ref{eq:F_gap}), where $T_{\rm L}' \equiv m(\ave{v_{1,j}^2}-\ave{v_{1,j}}^2) / k_{\rm B}$ and $T_{\rm R}' \equiv m(\ave{v_{N,j}^2}-\ave{v_{N,j}}^2) / k_{\rm B}$ are different from $T_{\rm L}$ and $T_{\rm R}$, respectively.
By using $F_{\rm gap}$ given in (\ref{eq:F_gap}),  we rewrite (\ref{eq:balance}) as 
\begin{equation}
-\left( \gamma_{\rm L} + \gamma_{\rm R} \right) S\ave{V}
+F_{\rm gap} + O(\epsilon^3)=0.
\label{eq:EOM} 
\end{equation}
It should be noted that $\ave{v_{i,j}}=\ave{V}=O(\epsilon)$.

%% Heat flux

We now derive the temperature gaps for the model we consider.
First, we shall find a relation connecting the temperature gap with the heat flux. 
Let $J_{{\rm L},j}$ be the heat flux transferred from the gas to the $j$-th solid particle at the left end, and $\dot{K}_{{\rm L},j}$ be the increasing rate of the kinetic energy per unit area of the $j$-th solid particle at the left end.
The energy conservation law leads to $J_{{\rm L},j}=\dot{K}_{{\rm L},j}-pv_{1,j}$.
Similarly, $J_{{\rm R},j} = \dot{K}_{{\rm R},j}+pv_{N,j}$ \cite{LDB}.
We denote by $\Delta K(v_{\rm G},v)$ the change in the kinetic energy of a solid particle for the collision of a solid particle of velocity $v$ with a gas particle of velocity $v_{\rm G}$, which is given  by $\Delta K(v_{\rm G},v) = 2m_{\rm G}m (v_{\rm G}-v)(m_{\rm G} v_{\rm G}+mv)/(m_{\rm G}+m)^2$.
We then calculate  $\ave{J_{{\rm L},j}}$ as $\ave{\int dv_{\rm G} \lambda_{\rm L}(v_{\rm G},v_{1,j}) \Delta K(v_{\rm G},v_{1,j})} -p\ave{v_{1,j}}$.
Expanding this expression in $\epsilon$, we obtain
\begin{eqnarray}
\ave{J_{{\rm L},j}} &=& \frac{\gamma_{\rm L}}{m} k_{\rm B}
\left( T_{\rm L} - T_{\rm L}'\right) + O(\epsilon^2), 
\label{eq:JL}
\\
\ave{J_{{\rm R},j}} &=& \frac{\gamma_{\rm R}}{m} k_{\rm B}
\left( T_{\rm R} - T_{\rm R}'\right) +O(\epsilon^2).
\label{eq:JR}
\end{eqnarray}
See Appendix \ref{sec:appendix} for the derivation.
These equations mean that the heat flux is related to the temperature gap \cite{Parrondo, Lepri}.
The average heat flux through the solid from the left to the right is written as $\ave{J}\equiv \ave{J_{{\rm L},j}} = -\ave{J_{{\rm R},j}}$ for any $j$.

%% Macroscopic Assumption and relate V to J

Second, we consider the heat flux.
In general, the heat flux depends on the interaction potential between solid particles, and it is difficult to calculate it from a microscopic description.
Nevertheless, by selecting a proper short-range interaction between solid particles in our model, we may phenomenologically assume Fourier's law in the form
\begin{equation}
\ave{J} = \kappa (T'_{\rm L}-T'_{\rm R}) / L, 
\label{eq:fourier}
\end{equation}
where $\kappa$ and $L$ represent the thermal conductivity and the axial length of the solid, respectively, and the temperature dependence of $\kappa$ is ignored. 
From (\ref{eq:JL}), (\ref{eq:JR}), and (\ref{eq:fourier}), we obtain
\begin{equation}
 \ave{J} = \frac{\kappa (T_{\rm L}-T_{\rm R})/L}
 {1+\kappa m ( 1/\gamma_{\rm L} + 1/\gamma_{\rm R} ) / (k_{\rm B}L) }
 + O(\epsilon^2),
 \label{eq:J_exp}
\end{equation}
and 
\begin{equation}
 T_{\rm L}' - T_{\rm L}= \frac{\gamma_{\rm R}/
 (\gamma_{\rm L}+\gamma_{\rm R})}
 {1+k_{\rm B}L\gamma_{\rm L}\gamma_{\rm R}/
 [ \kappa m(\gamma_{\rm L}+\gamma_{\rm R})]}
 (T_{\rm R}-T_{\rm L}) + O(\epsilon).
 \label{eq:Tgap_exp}
\end{equation}
By substituting this result into (\ref{eq:F_gap}), we obtain an expression for $F_{\rm gap}$ in terms of the experimental parameters.
Furthermore, (\ref{eq:F_gap}), (\ref{eq:EOM}),  (\ref{eq:JL}), and (\ref{eq:JR}) lead to the simple relation 
\begin{equation}
\ave{V} = -\frac{\pi\ave{J}}{8p} + O(\epsilon^2),
\label{eq:V-J}
\end{equation}
which connects the moving velocity with the heat flux passing inside the solid.
See Refs.~\cite{Fruleux,Kawai} for an intuitive explanation of the result.

%% Numerical experiment

\begin{figure}
\centering
\includegraphics[width=0.83\linewidth]{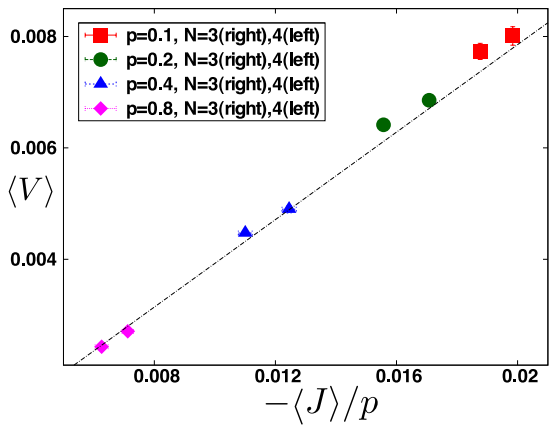}
\caption{ (Color online) Steady-state velocity $\ave{V}$ versus the heat flux over the pressure $-\ave{J}/p$ for $p=0.1, N=3,4$ (square, red), $p=0.2, N=3,4$ (circle, green), $p=0.4, N=3,4$ (triangle, blue), and $p=0.8, N=3,4$ (diamond, pink). The dotted line represents $\ave{V}=-\pi\ave{J}/(8p)$.}
\label{fig3}
\end{figure}

In order to directly demonstrate the validity of our theory, we performed numerical experiments by solving the Hamiltonian equations in combination with the Poisson process.
Here, for simplicity, we consider the case that the system is defined in two dimensions.
Concretely, we use the potentials $U_{\rm int}(\bm{r},\bm{r}')=k( \vert\bm{r}-\bm{r}'\vert -\sqrt{d} a)^2 /2$ for the $d$-th nearest neighbor pair of solid particles ($d=1,2$) and $U_{\rm w}(\bm{r})=1/\vert \bm{r}-\bm{r}_{\rm w}(\bm{r})\vert^6$, where $k$ is the spring constant, $\sqrt{d}a$ the natural length, and $\bm{r}_{\rm w}(\bm{r})$ the nearest position of the tube wall from $\bm{r}$.
All the quantities are converted into dimensionless forms by setting $k=a=m=1$.
We then set the parameter values as $k_{\rm B}T_{\rm L}=0.07$, $k_{\rm B}T_{\rm R}=0.1$, and $\epsilon = \sqrt{1/10}$.
Because the equations are nonlinear, we observed the temperature gradient inside the solid and the temperature gaps at the surface of the solid.
We then measured $\ave{V}$ and $\ave{J}$ for several values of $p$ and $N=M$, and plotted $(-\ave{J}/p, \ave{V})$ in Fig.~\ref{fig3}.
We find that the obtained data is consistent with the nontrivial relation (\ref{eq:V-J}).

\section{Experiments}
%% Experimental setup
Let us discuss the experimental feasibility of the phenomenon under consideration. 
As one example of laboratory experiments, we consider silicon and helium of atomic weight 28 and 4, respectively, where $\epsilon = \sqrt{1/7}$.
The thermal conductivity of silicon at room temperature is $\kappa \simeq 149\,\rm{J/(m\cdot s\cdot K)}$, and the density of silicon is  $\rho \simeq 2.33\,\rm{g/cm^3}$.
We set  $T_{\rm L}=293\,\rm{K}$, $T_{\rm R}=303\,\rm{K}$, $p= 1\,\rm{atm}$, $S=7\,\rm{cm^2}$, and $L=1\,\rm{cm}$.
By using (\ref{eq:F_gap}), (\ref{eq:J_exp}), (\ref{eq:Tgap_exp}), and (\ref{eq:V-J}), we obtain the velocity of the solid $\ave{V}\simeq 3.9\times 10\,{\rm cm/s}$, the temperature gap $T_{\rm L}'-T_{\rm L}\simeq 1.6\,{\rm K}$, and the temperature-gap-induced force $F_{\rm gap}\simeq 1.1\times 10^{-1}\,{\rm N}$.
These estimated values are large enough to be measured in careful experiments.
We next consider several possible difficulties that may arise in experiments.

%% Problem 1: friction

First, there is the friction between the solid and the tube.
Because the coefficient of static friction of a lubricant is at most $0.5$ \cite{lub}, the static friction force is about $0.5\times 9.8\,\rm{m/s^2}\times \rho SL \simeq 8.0\times 10^{-2}\,{\rm N}$, which is less than $F_{\rm gap}\simeq 1.1\times 10^{-1}\,\rm{N}$.
Thus, the effect of the friction can be mitigated by the use of a lubricant.

%% Problem 2: relaxation time

Second, one may worry that the relaxation time of the motion of the solid is longer than the observation time.
However, since the relaxation time is estimated as $mNM/(\gamma_{\rm L}+\gamma_{\rm R})S = \rho L / (\gamma_{\rm L}+\gamma_{\rm R}) \simeq 5.7\times 10^{-2}\,\rm{s}$, the velocity of the solid is rapidly relaxed to the steady-state value.

%% Problem 3: controll T_L and T_R

\begin{figure}
\centering
\includegraphics[width=0.82\linewidth]{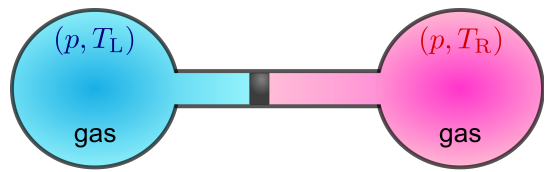}
\caption{(Color online) Simple realization of the phenomenon under consideration in an experiment. Large reservoirs of equilibrium gas are connected to the tube, and the length of the tube is assumed to be shorter than the mean free path of the gas particles. }
\label{fig4}
\end{figure}

\begin{figure}
 \centering
 \includegraphics[width=0.85\linewidth]{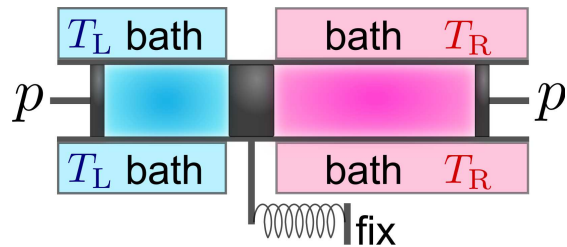}
\caption{(Color online) Two heat baths are in thermal contact with two dilute-gas regions, respectively. The solid material is immobilized by a spring. $F_{\rm gap}$ can be measured in the experiment, while the steady-state motion is not observed.}
\label{fig5}
\end{figure}

Lastly, the most difficult experimental setup may be the control of the temperatures of the gases.
One method is to connect the tube with baths of dilute gas, where the length of the tube is chosen to be shorter than the mean-free path of the dilute gas as shown in Fig.~\ref{fig4}.
The mean-free path of the helium atoms at $1\,\rm{atm}$ and $300\,\rm{K}$ is about $200\,\rm{nm}$ and is inversely proportional to the pressure.
Therefore, when the length of the tube is $20\,\rm{cm}$, we have to set the pressure at $1\times 10^{-6}\,\rm{atm}$, and $F_{\rm gap}$ becomes very small.
On the contrary, when the mean-free path is much shorter than the length of the tube, a temperature gradient appears in the gases.
This makes $T_{\rm R}-T_{\rm L}$ small, and as a result $F_{\rm gap}$ becomes small.
The simplest realization is to control the temperatures from the side wall of the tube.
In this case, we immobilize the solid by linking a spring to it (see Fig.~\ref{fig5}).
Because the value of the stall force is equal to $F_{\rm gap}$, we can measure its value, whereas we cannot observe the steady-state motion.

%% Application (Soret effect)

Setting aside the quantitative aspects, we may observe phenomena to which $F_{\rm gap}$ makes a dominant contribution.
One example is a Brownian particle under a temperature gradient \cite{Duhr, PiazzaParola, Sano, Wurger}.
If the thermal conductivity of the particle is much larger than that of a solution, heat flux passes inside the particle.
Then, because the temperature gap appears on the solid surface, the force $F_{\rm gap}$ is generated.
It should be noted, however, that other types of forces appear on the particle under a temperature gradient \cite{Wurger}.
It is a stimulating challenge to separate $F_{\rm gap}$ from the total force.

\section{Concluding remarks}
%% Summary
In this paper, we have predicted that the temperature gap at the interface between a solid and a gas yields the force $F_{\rm gap}$.
Because this force is not described by standard continuum theory such as hydrodynamics and elastic theory, further experimental and theoretical studies are necessary so as to obtain a systematic understanding of the nature of the force. 
Before ending the paper, we provide a few remarks.

%% Adiabatic piston

With regard to our setup, similar models were studied in the context of the so-called adiabatic piston problem \cite{Callen, Feynman, GruberLesne, Kestemont}.
Indeed, for the case $\kappa \gg Lk_{\rm B}\gamma_{\rm L}\gamma_{\rm R}/[m(\gamma_{\rm L}+\gamma_{\rm R})]$ or $N=1$, we obtain $\ave{J}=k_{\rm B} (T_{\rm L}-T_{\rm R})/[m(\gamma_{\rm L}^{-1}+\gamma_{\rm R}^{-1})] + O(\epsilon^2)$ and $\ave{V}=\sqrt{\pi/8}\epsilon (\sqrt{k_{\rm B}T_{\rm R}/m}-\sqrt{k_{\rm B}T_{\rm L}/m})+O(\epsilon^2)$, where $T_{\rm L}' = T_{\rm R}'$.
These expressions for $\ave{J}$ and $\ave{V}$ are identical to those derived in Refs. \cite{GruberPiasecki, GruberFrachebourg}.
Here, it should be noted that $m$ in the preceding studies was assumed to be the total mass of the solid, which gives much smaller values of $\ave{J}$ and $\ave{V}$ than ours. 

%% Future work (macroscopic equations)

In contrast to the interface case, there exists no force due to temperature differences at the atomic scale in the bulk.
In order to clearly understand the difference between the two cases, we should derive $F_{\rm gap}$ on the basis of a mechanical description of solids and gases.
This fundamental question might be solved by considering hydrodynamics of a binary mixture fluid in the phase separation state.
Because hydrodynamics may involve the discontinuity of the temperature profile at the interface between the two materials, the standard assumption of slowly varying thermodynamic quantities may not be valid.
The derivation may be obtained as an extension of a recent work \cite{Sasa} in which the hydrodynamic equations for a simple fluid are derived from a Hamiltonian description of identical particles.
Obviously, the experimental measurement of $F_{\rm gap}$ is of great importance even for the theory.
By clarifying the mechanism of nonstandard forces, we hope to develop the understanding of nonequilibrium systems.

\begin{acknowledgments}
The authors express special thanks to Y. Nakayama for continuous discussions, especially on the experimental feasibility of the phenomenon under consideration.
They also thank K. Sekimoto, R. Kawai, K. Saito, K. Kawaguchi, K. Takeuchi, M. Sano, K. Kanazawa, T. Nemoto, and N. Shiraishi for useful discussions.
The present study was supported by KAKENHI, Grants No. 22340109, No. 25103002, and by the JSPS Core-to-Core program ``Non-equilibrium dynamics of soft-matter and information.''
\end{acknowledgments}

\appendix
\section{Derivation of (\ref{eq:balance}), (\ref{eq:JL}), and (\ref{eq:JR})} \label{sec:appendix}
By considering $f_{\rm eq}^{\rm L}(v)=\epsilon \sqrt{\beta_{\rm L}m/2\pi}\exp (-\epsilon^2 \beta_{\rm L}mv^2 /2)$ and $f_{\rm eq}^{\rm R}(v)=\epsilon \sqrt{\beta_{\rm R}m/2\pi}\exp (-\epsilon^2 \beta_{\rm R}mv^2 /2)$, we obtain
\begin{align}
 \int_{V}^{\infty}v^k f_{\rm eq}^{\rm L}(v) dv
 &= \int_{0}^{\infty}v^k f_{\rm eq}^{\rm L}(v)dv + O(\epsilon)
 \notag
 \\
 & = \frac{\epsilon^{-k}}{2\sqrt{\pi}} \left( \frac{2}{\beta_{\rm L} m}\right)^{\frac{k}{2}} \Gamma \left( \frac{k+1}{2}\right) + O(\epsilon),
 \label{eq:key1}
 \\
 \int_{-\infty}^{V}v^k f_{\rm eq}^{\rm R}(v)dv
 &= \int_{-\infty}^{0}v^k f_{\rm eq}^{\rm R}(v)dv + O(\epsilon)
 \notag
 \\
 &= \frac{(-\epsilon)^{-k}}{2\sqrt{\pi}} \left( \frac{2}{\beta_{\rm R} m}\right)^{\frac{k}{2}} \Gamma \left( \frac{k+1}{2}\right) + O(\epsilon),
 \label{eq:key2}
\end{align}
where $k$ is a non-negative integer.
By using (\ref{eq:key1}) and (\ref{eq:key2}), we obtain
\begin{align}
 &\int \lambda_{\rm L}(v, V) I(v,V) dv
 \notag
 \\
 &\qquad = \frac{2\epsilon^2 mn_{\rm L}}{1+\epsilon^2}
 \int_{V}^{\infty}(v-V)^2 f_{\rm eq}^{\rm L}(v) dv
 \notag
 \\
 &\qquad = (1-\epsilon^2) p -\gamma_{\rm L} V
 + \epsilon^2 p\beta_{\rm L} mV^2 + O(\epsilon^3),
 \label{eq:I_L}
 \\
 &\int dv\; \lambda_{\rm R}(v, V) I(v,V)
 \notag
 \\
 &\qquad = -\frac{2\epsilon^2 mn_{\rm R}}{1+\epsilon^2}
 \int_{-\infty}^{V} dv (V-v)^2 f_{\rm eq}^{\rm R}(v)
 \notag
 \\
 &\qquad = - (1-\epsilon^2)p -\gamma_{\rm R} V
 - \epsilon^2 p\beta_{\rm R} mV^2 + O(\epsilon^3),
 \label{eq:I_R}
 \\
 &\int \lambda_{\rm L}(v,V) \Delta K(v,V)dv
 \notag
 \\
 &\qquad = \frac{2\epsilon^2 mn_{\rm L}}{(1+\epsilon^2)^2}
 \int_{V}^{\infty} (v-V)^2 (\epsilon^2 v + V) f_{\rm eq}^{\rm L}(v)dv
 \notag
 \\
 &\qquad = pV - \gamma_{\rm L} V^2 + \frac{\gamma_{\rm L}k_{\rm B}T_{\rm L}}{m} + O(\epsilon^2),
 \label{eq:K_L}
 \\
 &\int \lambda_{\rm R}(v,V) \Delta K(v,V)dv
 \notag
 \\
 &\qquad = -\frac{2\epsilon^2 mn_{\rm R}}{(1+\epsilon^2)^2}
 \int_{-\infty}^{V} (V-v)^2 (\epsilon^2 v + V) f_{\rm eq}^{\rm R}(v)dv
 \notag
 \\
 &\qquad = -pV - \gamma_{\rm R} V^2 + \frac{\gamma_{\rm R}k_{\rm B}T_{\rm R}}{m} + O(\epsilon^2),
 \label{eq:K_R}
\end{align}
where we have used $p=n_{\rm L}/\beta_{\rm L}=n_{\rm R}/\beta_{\rm R}$, $\gamma_{\rm L}=\epsilon n_{\rm L}\sqrt{8mk_{\rm B}T_{\rm L}/\pi}$, and $\gamma_{\rm R}=\epsilon n_{\rm R}\sqrt{8mk_{\rm B}T_{\rm R}/\pi}$.
Here, we assume that $\ave{O(\epsilon^k)} = O(\epsilon^k)$.
Then, (\ref{eq:I_L}) and (\ref{eq:I_R}) lead to
\begin{align}
 &\sum_{j=1}^{M} \Big[ \ave{F_{{\rm L},j}} + \ave{F_{{\rm R},j} } \Big]
 \notag
 \\
 &\quad = \sum_{j=1}^{M} \left[ \ave{\int dv_{\rm G}\lambda_{\rm L}(v_{\rm G},v_{1,j})I(v_{\rm G},v_{1,j})} \right.
 \notag
 \\
 &\qquad \quad \left. + \ave{\int dv_{\rm G}\lambda_{\rm R}(v_{\rm G},v_{N,j})I(v_{\rm G},v_{N,j})} \right] \frac{S}{M}
 \notag
 \\
 &\quad = \sum_{j=1}^{M} \left[ -\gamma_{\rm L}\ave{v_{1,j}}
 -\gamma_{\rm R}\ave{v_{N,j}}
 + \epsilon^2 p\beta_{\rm L}m \ave{v_{1,j}^2}\right.
 \notag
 \\
 &\qquad \quad \left. - \epsilon^2 p\beta_{\rm R}m \ave{v_{N,j}^2}\right]
 \frac{S}{M}+O(\epsilon^3).
 \label{eq:EOM_pre}
\end{align}
By using $\ave{v_{i,j}}=\ave{V}=O(\epsilon)$, (\ref{eq:K_L}), and (\ref{eq:K_R}), we obtain
\begin{align}
 \ave{J_{{\rm L},j}}
 &= \ave{\int dv_{\rm G} \lambda_{\rm L}(v_{\rm G},v_{1,j})
 \Delta K(v_{\rm G},v_{1,j})} -p\ave{v_{1,j}}
 \notag
 \\[3pt]
 & = \frac{\gamma_{\rm L}}{m}k_{\rm B}\left( T_{\rm L} -
 m\ave{v_{1,j}^2}/k_{\rm B} \right) + O(\epsilon^2)
 \notag
 \\[3pt]
 & = \frac{\gamma_{\rm L}}{m}k_{\rm B}
 \left( T_{\rm L} - T_{\rm L}'\right) + O(\epsilon^2),
 \\[3pt]
 \ave{J_{{\rm R},j}}
 &= \ave{\int dv_{\rm G} \lambda_{\rm R}(v_{\rm G},v_{N,j})
 \Delta K(v_{\rm G},v_{N,j})} +p\ave{v_{N,j}}
 \notag
 \\[3pt]
 & = \frac{\gamma_{\rm R}}{m}k_{\rm B}\left( T_{\rm R} -
 m\ave{v_{N,j}^2}/k_{\rm B} \right) + O(\epsilon^2)
 \notag
 \\[3pt]
 & = \frac{\gamma_{\rm R}}{m}k_{\rm B}
 \left( T_{\rm R} - T_{\rm R}'\right) + O(\epsilon^2).
\end{align}

%%%%%%%%%%%%%%%%%%%%%%%
% References          %
%%%%%%%%%%%%%%%%%%%%%%%


\begin{thebibliography}{99}
\bibitem{LDB} It can be proved that $J_{{\rm L},j}$ and $J_{{\rm R},j}$ satisfy the local detailed balance condition.
\end{thebibliography}
\end{document}